\documentclass[a4paper,11pt]{article}
\usepackage{pos}
\usepackage{natbib}
\usepackage{upgreek}
\graphicspath{{logos/},{fig/}}

\title{The design and expected performance of the ALICE ITS3 upgrade}
\ShortTitle{Design and expected performance of the ALICE ITS3}

\author*[a]{J. Sonneveld for the ALICE Collaboration}

\affiliation[a]{Nikhef,\\
  Science Park 105, Amsterdam, the Netherlands}

  \emailAdd{jory.sonneveld@nikhef.nl}

\abstract{
During the LHC Long Shutdown 3 (2026--29) ALICE will replace its three innermost tracking layers by a new detector, the “ITS3”. It will be based on newly developed, wafer-scale monolithic active pixel sensors, which are bent into truly cylindrical layers and held in place by light mechanics made from carbon foam. Unprecedented low values of material budget (0.09\% $X_0$ per layer) and proximity to interaction point (19 mm) lead to a factor two improvement in pointing resolutions for particles from very low $p_{\mathrm{T}}$ ($\mathcal{O}(100$ MeV/$c$)), achieving, for example, 20$~\upmu$m and 15$~\upmu$m in the transversal and longitudinal directions, respectively, for 1 GeV/$c$ particles. After a successful R\&D phase (2019--2023), which demonstrated the feasibility of this innovative detector and led to the Technical Design Report \cite{its3tdr}, the final sensor and mechanics are being developed right now. This contribution will review the conceptual design and the main R\&D achievements, as well as the current activities and road to completion and installation. It includes a projection of the improved physics performance, in particular for heavy-flavor mesons and baryons, as well as for thermal dielectrons that will come into reach with this new detector installed.
}

\FullConference{The European Physical Society Conference on High Energy Physics (EPS-HEP2025)\\
7-11 July 2025\\
Marseille, France\\}

\begin{document}
\maketitle

\section{The ALICE Inner Tracking System Upgrade}
The ALICE experiment \cite{alice} at the Large Hadron Collider \cite{lhc} at CERN is the first to use monolithic active pixel sensors at this most powerful collider in the world. The inner three thin layers are only 0.36\% of $X_0$ and the innermost layer is only 22~mm away from the beam line, the third is 42 mm from the beam line. The outer layers at 194--395~mm from the beam line have a mere 1.1\% of $X_0$. The ALICE PIxel DEtector (ALPIDE) sensors are 3~cm$~\times~$1.5~cm with 27$~\upmu$m~$\times$~29$~\upmu$m pixels and are only 50$~\upmu$m thick in the innermost three layers. With 12.5 billion pixels and 10$~$m$^2$ active area, the current Inner Tracking System, or ITS2, is the largest pixel detector ever built \cite{its2tdr, its2ls2}. It has been successfully taking data since September 2021.

\begin{figure}
    \centering
    \begin{minipage}{0.5\textwidth}
\includegraphics[width=0.47\linewidth]{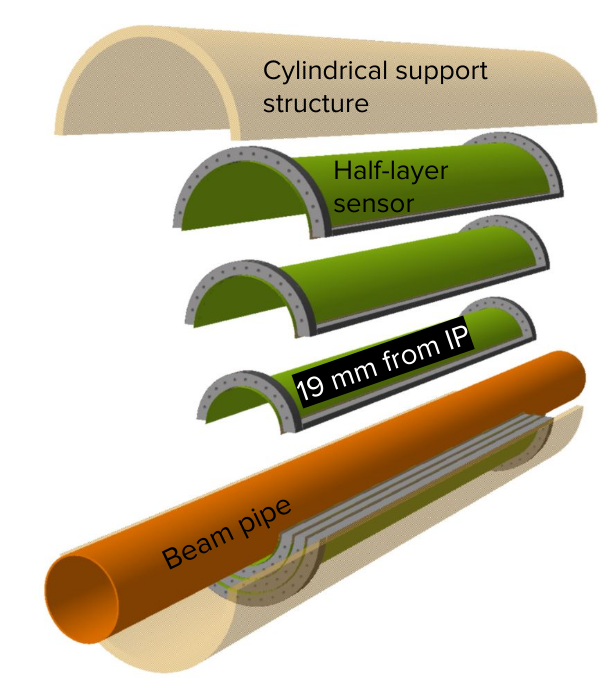}   \includegraphics[width=0.47\linewidth]{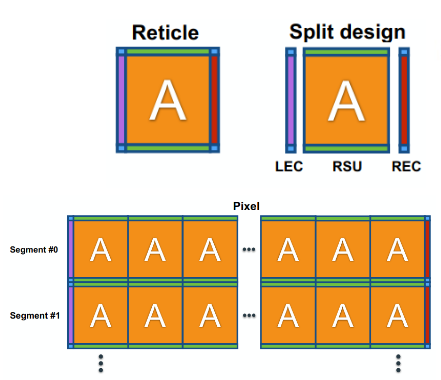}
    \caption{Left: The ALICE ITS3 upgrade will have three layers. Each layer consists of two stitched and bent 27 cm-long sensors. Right: The process of stitching allows for a repeated sensor unit (RSU) and a left endcap (LEC) and a right endcap (REC) to be made into one wafer-scale sensor. Figures are taken from \cite{its3tdr}.}
    \label{fig:its3drawing}
    \label{fig:stitching}
        \end{minipage}
        \hspace{0.02\textwidth}
        \begin{minipage}{0.46\textwidth} 
         \includegraphics[width=\linewidth]{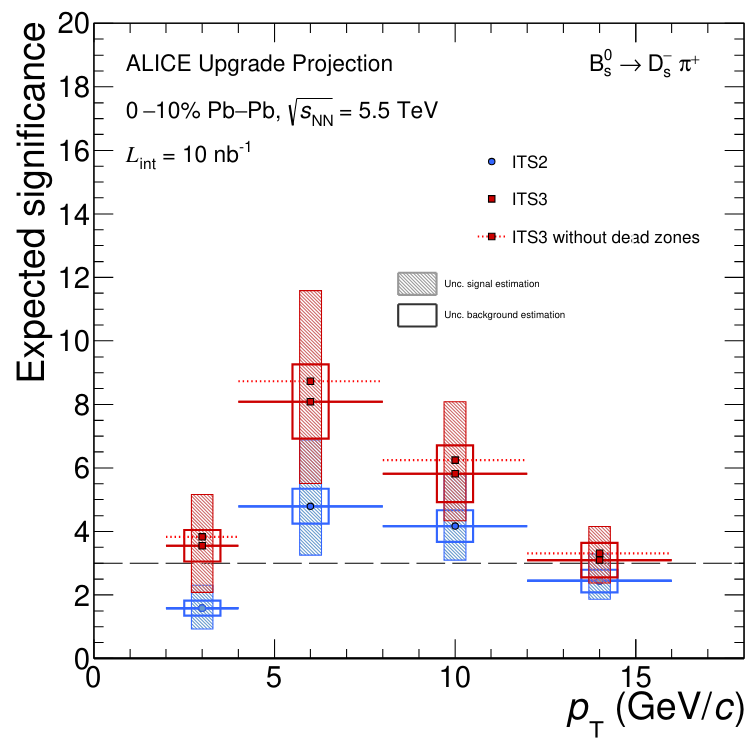}
    \caption{The ITS3 will allow for measurements of B-meson decays with much better precision than the ITS2 today. Figure taken from Ref. \cite{its3tdr}.}
    \label{fig:bs}
            \end{minipage}
\end{figure}

In Run 4, ALICE aims to focus, among other things, on precise detection of forward photons to probe small-$x$ initial gluons;
low-$p_{\mathrm{T}}$ heavy-flavor hadrons for studying the heavy-quark thermalization;
and low-mass dileptons for determining the Quark--Gluon Plasma (QGP) temperature.
For this, ALICE needs precise vertexing with a high impact-parameter resolution, and a pointing resolution of the order of 100 $\upmu$m for $\sim$100 MeV/$c$ charged particles. In 2028, a replacement of the inner barrel of ITS2 is foreseen, where the 432 chips of the innermost three layers will be replaced by the ALICE Inner Tracking System 3 (ITS3) with 6 stitched, bent wafer-scale sensors made of 300 mm wafers (see Fig. \ref{fig:its3drawing}) \cite{its3tdr}. This reduces the material budget from 0.36\% of $X_0$ to a mere 0.09\% of $X_0$. The innermost layer will move from 22.4 mm from the interaction point (IP) to only 19 mm from the IP. To achieve this, a new beam pipe having a 500-$\upmu$m-thick wall and with only 16.5 mm radius will replace the current beam pipe that has an 18 mm radius and a 800-$\upmu$m-thick wall, reducing the material budget of the beam pipe from 0.22\% to only 0.14\% of $X_0$.

The process of stitching enables to interconnect neighboring design blocks during the CMOS lithographic process, and as such one can make a chip larger than the field of view of the lithographic equipment. For the ITS3, a split design allows for a repeated stitched unit and a left and right end cap at the end of the wafer (see Fig. \ref{fig:stitching}). The 6 half-layer sensors planned for the ITS3 will have $22.8~\upmu$m$~\times ~20.8~\upmu$m pixels, will be thinned to 50$~\upmu$m, and will be mechanically held in place by carbon-foam spacers. The innermost layer of the detector, only 19~mm from the LHC beam, will be exposed to fluences of $\Phi_{\mathrm{eq}} = 4 \times 10^{12}~1~{\rm MeV} \mathrm{n_{eq}}/\mathrm{cm}^2$ and a dose of 4~kGy, and will have to cope with interaction rates of 2.2~MHz/cm$^2$.

\section{Improved measurements with precise tracking and vertexing}
\begin{figure}
    \centering
    \includegraphics[width=\linewidth]{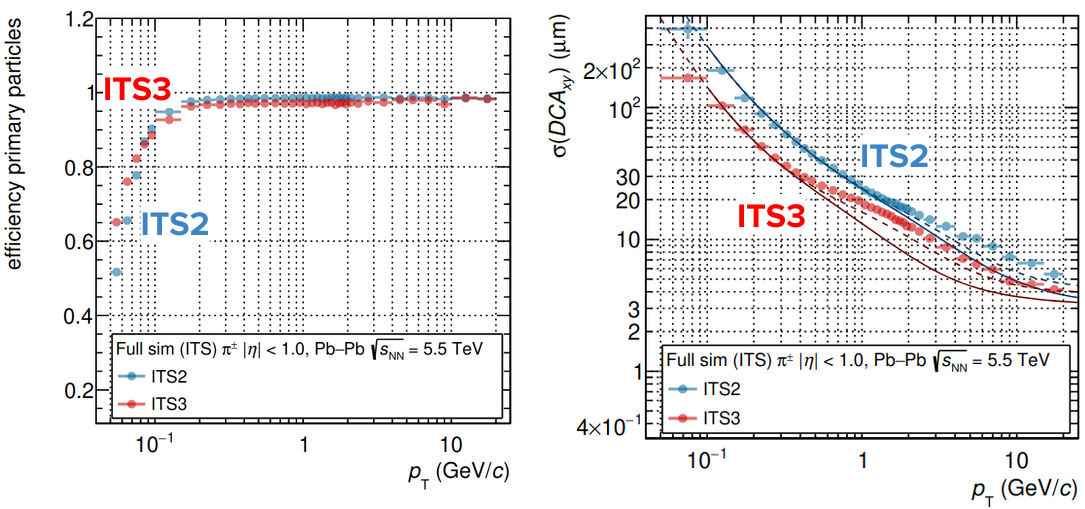}
    \caption{Tracking efficiency and impact parameter resolution for primary particles in Pb-Pb collisions as measured using the present ITS2 and the ITS2 with the ITS3 layers. Especially the pointing resolution improves by a factor 2 over almost all momenta. Figure taken from Ref. \cite{its3tdr}.}
    \label{fig:trackingvertexing}
\end{figure}
\begin{figure}
    \centering
\includegraphics[width=\linewidth]{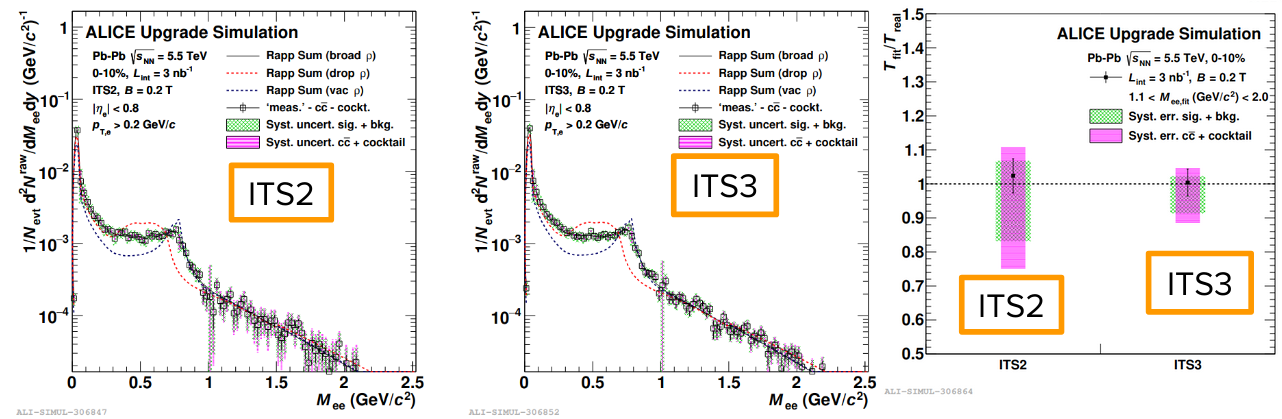}
    \caption{The QGP temperature can be measured from thermal dielectrons. The ITS3 reduces the systematic error on the temperature by a factor two compared with what the ITS2 can achieve today. Figure taken from Ref. \cite{its3tdr}.}
    \label{fig:qgptemp}
\end{figure}

The ITS3 will have an improved tracking and vertexing performance compared to the current ITS2 that is in operation in ALICE. At low transverse momenta there will be an improved detection of primary particles. At the same time, the pointing resolution, which is the error on the distance of closest approach in the $x-y$ plane, is expected to improve by a factor 2 over almost all momenta, as shown in Fig.~\ref{fig:trackingvertexing}. This improves the significance of certain measurements. For example, CMS made a first measurement of the ratio of $\mathrm{B}_{\mathrm{s}}^0 /\mathrm{B}^+$ yield ratio in Pb--Pb collisions and pp collisions, albeit with large uncertainties \cite{CMSb_to_bs}. ALICE similarly measured non-prompt D mesons \cite{aliceds}.
Both ALICE and CMS see an enhancement in this ratio, but neither made a significant observation. The ITS3 can measure at much lower transverse momentum, and will increase the significance of the measurement as shown in Fig. \ref{fig:bs}, all thanks to a very low material budget. The ITS3 can also contribute to a much more precise measurement of the QGP temperature reducing the systematic error by a factor of 2, see Fig. \ref{fig:qgptemp}. This is thanks to the high-quality low-$p_{\mathrm{T}}$ tracking of ITS3, resulting in improved photon-conversion reconstruction efficiency as well as improved vertexing performance and reduced backgrounds for low-mass di-electrons.

\section{How to reduce the material?}
\begin{figure}
     \centering
\includegraphics[width=0.8\linewidth]{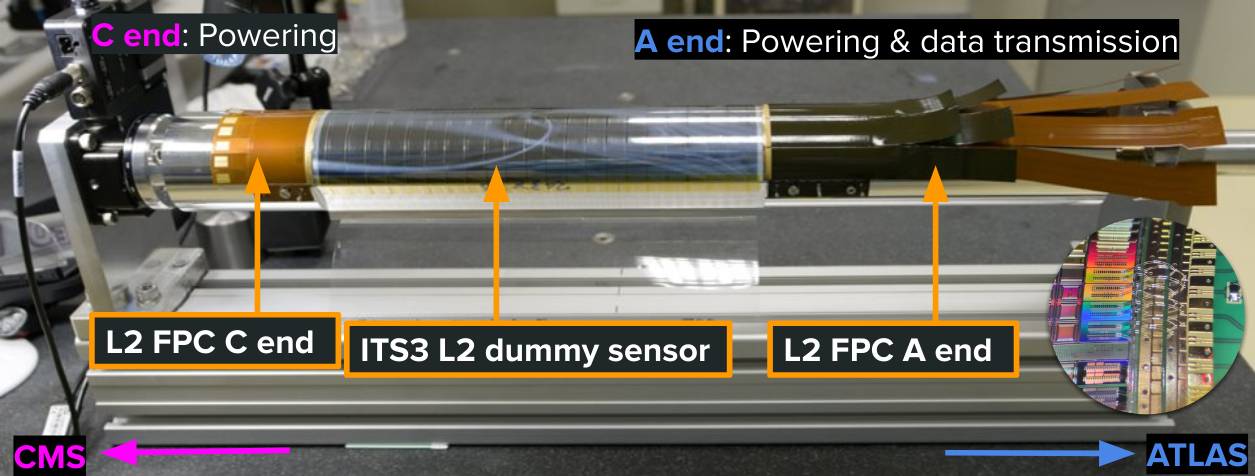}
     \caption{Full-scale engineering model of an ITS3 detector including a dummy sensor and flexible PCBs for powering and readout. The wire bonding of such a bent sensor, shown in the figure, needs special care and tools.}
     \label{fig:engineeringmodel}
 \end{figure}

To achieve a low material budget in the ITS3, a number of materials in the ITS2 
will be removed or replaced. The bent sensors will be held in place by carbon-foam spacers and will be cooled using air cooling. The cooling plate and space frame of ITS2 are then no longer needed. This requires to keep the sensor power below $\sim40$~mW/cm$^2$. As the long sensors will be connected to a PCB only at the ends, the glue dots and PCBs along the previous ITS2 stave will also be left out.
 This can be seen from the full-scale engineering model in Fig. \ref{fig:engineeringmodel}, where a flexible PCB is connected to each end of the sensor-long detector but is not placed on top of the sensor. One end provides powering only, the other end provides both powering and readout.

\begin{figure}
    \centering
    \includegraphics[width=\textwidth]{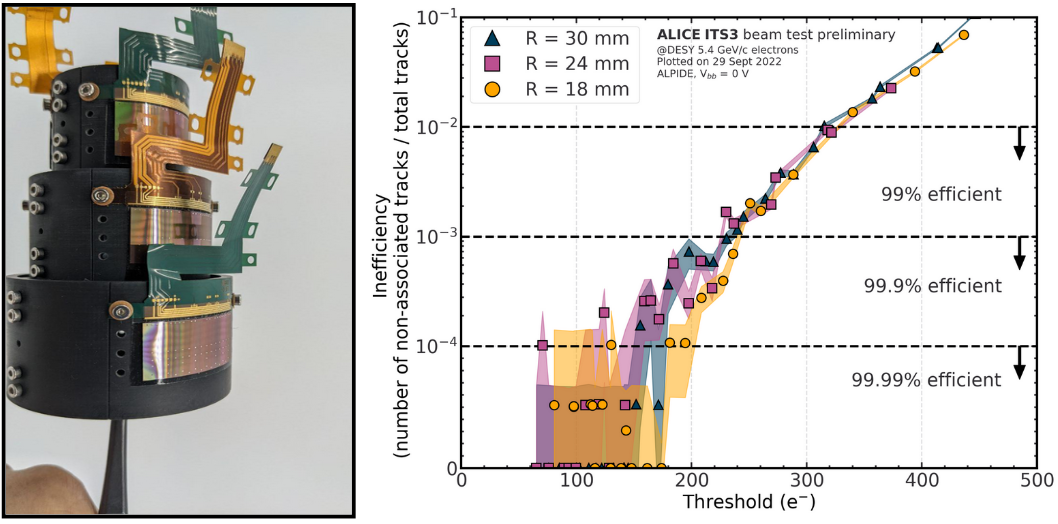}
    \caption{A full mockup of the final ITS3 called the $\upmu$ITS3 (left) is shown to be fully efficient with ALPIDEs bent to the three ITS3 radii (right). Figures from Ref. \cite{bentalpides}.}
    \label{fig:bentalpides}
\end{figure}

\begin{figure}
    \centering
    \includegraphics[width=\linewidth]{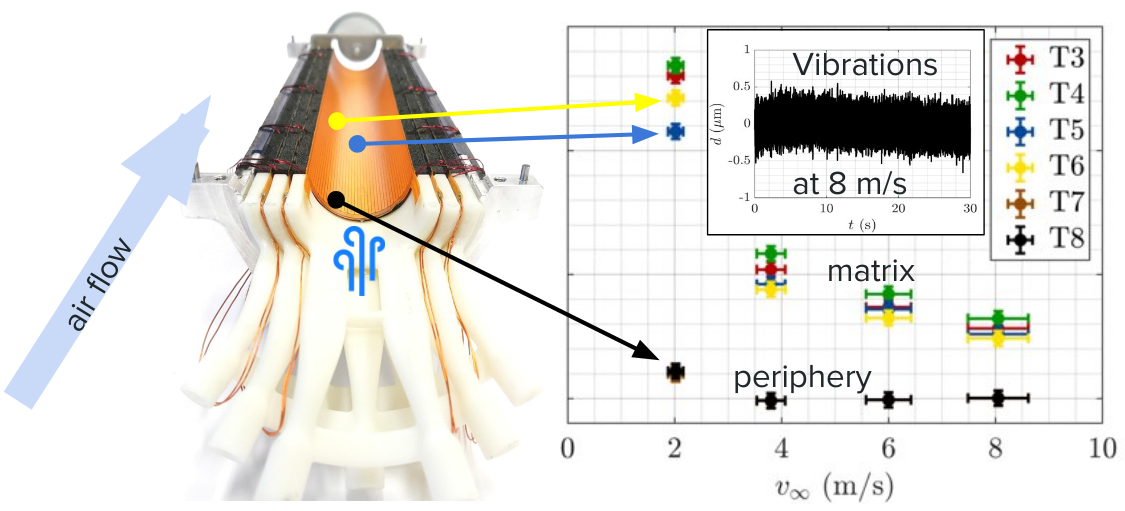}
    \caption{Air cooling has been demonstrated with a mockup where at 8 m/s vibrations are within 1 $\upmu$m. The colored markers are for sensors in the pixel matrix area, the black markers in the periphery of the chip. The inset shows the vibrations under air cooling at 8 m/s. Figures taken from \cite{its3tdr}.}
    \label{fig:aircooling}
\end{figure}

The bending of silicon wafers and functional ALPIDEs is now routine in ALICE. A full mock-up called the  “$\upmu$ITS3” was bent to ITS3 radii and tested in charged particle beams. The spatial resolution was found to be uniform among different radii, and the efficiency and resolution consistent with flat ALPIDEs \cite{bentalpides, bentalpides2}. This is also shown in Fig. \ref{fig:bentalpides}.
A model for thermal studies with temperature sensors and heaters has demonstrated that air cooling at 8 m/s can limit the temperature gradient to a few degrees over the length of the sensor, and that vibrations perpendicular to the beam line are within 1 $\upmu$m peak to peak. This is also shown in Fig. \ref{fig:aircooling}.

\section{Sensor R\&D}
\begin{figure}
    \centering
\includegraphics[width=\linewidth, trim=0 0 537 0, clip]{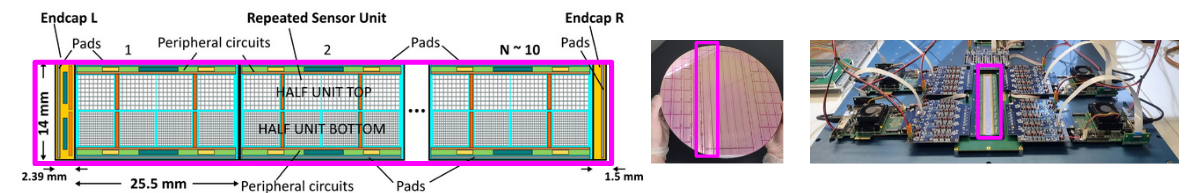}\\
\includegraphics[width=\linewidth, trim=643 0 0 0, clip]{figs/MOSS.png}
    \caption{The MOSS from Engineering Run 1. Top: Schematic of the MOSS with a repeated sensor unit (RSU) and a left (LEC) and a right endcap (REC) for powering and readout. Bottom left: A wafer with MOSS chips. Bottom right: A MOSS test system in the laboratory.}
    \label{fig:MOSS}
\end{figure}

Monolithic active pixel sensors with small collection electrodes have the advantage of reduced material budget as electronics is integrated into the sensor, a low capacitance, and low noise. For a detector designed to be as large as the ITS3, the TPSCo CIS 65-nm technology was chosen as it is produced on 300 mm wafers allowing for 26-cm-long sensors. In Engineering Run 1, the first stitched sensor prototypes were produced: the Monolithic Stitched Sensor (MOSS) and with Timing (MOST): a picture is shown in Fig. \ref{fig:MOSS}.  
    The MOSS covers an area of $14~ \times ~259~$mm$^2$ with 6.72 million pixels, and the MOST $2.5 ~\times ~259~$mm$^2$ with 0.9 million pixels, with $22.5~\times ~22.5~\upmu$m$^2$ pixels and $18~\times ~18~\upmu$m$^2$, respectively. A processing issue in ER1 caused shorts in metal layers \cite{metalshort}, which is now understood by the foundry and fixed for the next submission. The yield in unaffected MOSSs is larger than $90 \%$, showing that these are failure-tolerant structures. With analog pixel test structures, the 65 nm technology has been validated to ITS3 neutron fluences \cite{apts} and to have a time resolution of 67 ps \cite{aptsoa}. The digital test structure has also been validated to neutron fluences beyond the ITS3 \cite{dpts}. The 259-mm-long stitched sensors have also been shown to be fully functional even after irradiation, and meet the ITS3 requirements at neutron fluences of $\Phi_{\mathrm{eq}} = 4 \times 10^{12}~1~\mathrm{MeV}~\mathrm{n}_{\mathrm{eq}}/\mathrm{cm}^2$ and a dose of 400 krad \cite{moss}. This was found with data collected during 82 days of in-beam data taking with 17 chips of 2 different sensor-process variations \cite{processmod} operated at many different settings, and irradiated to 5 different irradiation levels. Irradiations were conducted with neutrons up to 30~MeV in Ljubljana and 10~keV electrons at CERN.

\begin{figure}
    \centering
    \includegraphics[width=\linewidth]{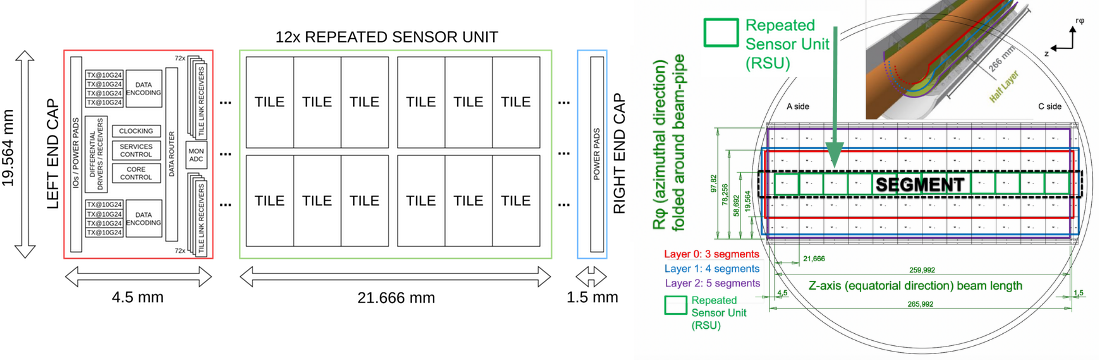}
    \caption{The MOSAIX sensor will have tiles with independent powering, control and readout. It has two endcaps like the current prototypes. Figure taken from Ref. \cite{its3tdr}.}
    \label{fig:mosaix}
\end{figure}

The next stitched prototype to be produced in R\&D towards ITS3 in Engineering Run 2 will be the MOnolithic Stitched Active pIXel, or MOSAIX for short, which is foreseen to be the final prototype before the ITS3 production \cite{mosaix}. It is to be tested with bending and includes features from both the MOSS and MOST. It has a modular design, with 3, 4, or 5 segments for each of the layers 0--2, as shown in the right panel of Fig. \ref{fig:mosaix}. Readout is through the left endcap and powering through both the left and right endcap. The sensor has 12 repeated sensor units (RSUs), with 12 tiles per RSU. Each tile has independent powering, control, and readout. In total this amounts to 144 tiles, each covering 0.7\% of the MOSAIX area. These can be switched off individually in case of a short or issue. The sensitive area of the sensor is in total 93\%. A schematic of the MOSAIX is shown in the left panel of Fig.~\ref{fig:mosaix}.

\section{Outlook}
In summary, ALICE will install new inner layers for the Inner Tracking System 3 (ITS3) for the LHC Run 4 in 2030. ALICE aims for a detector with truly cylindrical, wafer-scale monolithic active pixel sensors. Silicon flexibility and bending has been proven with routine bending tests, and a full mockup of the ITS3 has been shown to be efficient when bent to the ITS3 target radii. Sensor prototypes reach more than 99\% detection efficiency and less than 
10$^{-6}$ hits/pixel/event fake-hit rate at room temperature at the ITS3 neutron fluence requirement of $\Phi_{\mathrm{eq}} = 4 \times 10^{12}~1~\mathrm{MeV}~\mathrm{n}_{\mathrm{eq}}/\mathrm{cm}^2$. The first stitched sensors have been successfully tested, and the final prototype is coming next year. The ITS3 R\&D will pave the way to 50-$\upmu$m-thin, low-power ($\sim$40 mW/cm$^2$) sensors that could be used in ALICE~3 (see Refs. \cite{alice3}) and beyond at the FCC. The ITS3 is a successful R\&D project enabling a wealth of new precision measurements.


\begin{thebibliography}{99}
\bibitem{alice}
ALICE Collaboration, “The ALICE experiment at the CERN LHC”, \textit{JINST} \textbf{3} (2008) S08002, \href{https://doi.org/10.1088/1748-0221/3/08/S08002}{doi:10.1088/1748-0221/3/08/S08002}.

\bibitem{lhc}
Oliver Sim Brüning, Paul Collier, P Lebrun, Stephen Myers, Ranko Ostojic, John Poole, and Paul Proudlock. “LHC Design Report.” \textit{CERN Yellow Reports: Monographs} 2004. \href{https://doi.org/10.5170/CERN-2004-003-V-1}{doi:10.5170/CERN-2004-003-V-1}.

\bibitem{its2tdr}
ALICE Collaboration, “Technical design report for the upgrade of the ALICE inner tracking system" \textit{J. Phys. G: Nucl. Part. Phys.} \textbf{41} (2014) 087002, \href{https://doi.org/10.1088/0954-3899/41/8/087002}{doi:10.1088/0954-3899/41/8/087002}.

\bibitem{its2ls2}
ALICE Collaboration, “ALICE upgrades during the LHC Long Shutdown 2”, \textit{JINST} \textbf{19} (2024) P05062,
\href{https://doi.org/10.1088/1748-0221/19/05/P05062}{doi:10.1088/1748-0221/19/05/P05062}

\bibitem{its3tdr}
ALICE collaboration. “Technical Design report for the ALICE Inner Tracking System 3-ITS3; A bent wafer-scale monolithic pixel detector”, CERN-LHCC-2024-003, ALICE-TDR-021 (2024), \href{https://cds.cern.ch/record/2890181/}{cds:2890181}.

\bibitem{CMSb_to_bs}
CMS Collaboration, "Observation of $\mathrm{B}_{\mathrm{s}}^0$ mesons and measurement of the $\mathrm{B}_{\mathrm{s}^0} /\mathrm{B}^+$ yield ratio in PbPb collisions at $\sqrt{s_{\mathrm{NN}}} = 5.02~$TeV", \textit{Phys. Lett. B} \textbf{829} (2022) 137062, \href{https://doi.org/10.1016/j.physletb.2022.137062}{doi:10.1016/j.physletb.2022.137062}



\bibitem{aliceds}
ALICE Collaboration, "Measurement of beauty-strange meson production in Pb–Pb collisions at $\sqrt{s_{\mathrm{NN}}} = 5.02~$TeV via non-prompt $\mathrm{D}_{\mathrm{s}}^+$ mesons",  \textit{Phys. Lett. B} \textbf{846} (2023) 137561, \href{https://doi.org/10.1016/j.physletb.2022.137561}{doi:10.1016/j.physletb.2022.137561}

\bibitem{bentalpides}
ALICE ITS Project, "First demonstration of in-beam performance of bent Monolithic Active Pixel Sensors", \textit{Nucl. Inst. M. A} \textbf{1028} (2022) 166280, \href{https://doi.org/10.1016/j.nima.2021.166280}{doi:10.1016/j.nima.2021.166280}.

\bibitem{bentalpides2}
A. Andronic et. al., "Detection efficiency and spatial resolution of Monolithic Active Pixel Sensors bent to different radii", \href{https://doi.org/10.48550/arXiv.2502.04941}{arXiv:2502.04941} (2025).
%


\bibitem{moss}
O. Abdelrahman \textit{et al.}, "Characterisation of the first wafer-scale prototype for the ALICE ITS3 upgrade: the monolithic stitched sensor (MOSS)", 2025, \href{https://doi.org/10.48550/arXiv.2510.11463}{doi:10.48550/arXiv.2510.11463}

\bibitem{metalshort}
G. Aglieri Rinella, et al., In submission to IEEE Trans. Nucl. Sci. (2025).

\bibitem{apts}
G. Aglieri Rinella \textit{et al.}, "Characterisation of analogue Monolithic Active Pixel Sensor test structures implemented in a 65 nm CMOS imaging process," \textit{Nucl. Instr. Meth. A} \textbf{1069} 169896 (2024) doi:\href{https://doi.org/10.1016/j.nima.2024.169896}{10.1016/j.nima.2024.169896}.

\bibitem{aptsoa}
G. Aglieri Rinella et al., “Time performance of Analog Pixel Test Structures with in-chip operational amplifier implemented in 65 nm CMOS imaging process”, \textit{Nucl. Instr. Meth. A}, (2024) \textbf{1070} 170034, \href{https://doi.org/10.1016/j.nima.2024.170034}{doi:10.1016/j.nima.2024.170034}.

\bibitem{dpts}
G. Aglieri Rinella \textit{et al.}, "Digital pixel test structures implemented in a 65 nm CMOS process," \textit{Nucl. Instr. Meth. A} \textbf{1056} (2023) 168589, doi:\href{https://doi.org/10.1016/j.nima.2023.168589}{10.1016/j.nima.2023.168589}



\bibitem{processmod}
W. Snoeys \textit{et al.}, "A process modification for CMOS monolithic active pixel sensors for enhanced depletion, timing performance and radiation tolerance," \textit{Nucl. Instr. Meth. A}, \textbf{871} (2017) 90-96, doi:\href{https://doi.org/10.1016/j.nima.2017.07.046}{10.1016/j.nima.2017.07.046}

\bibitem{mosaix}
P. Vicente Leitao on behalf of the ALICE Collaboration, "Development of the MOSAIX chip for the ALICE ITS3 upgrade," \href{http://doi.org/10.1088/1748-0221/20/06/C06001}{\textit{JINST} \textbf{20} (2025) C06001}.


\bibitem{alice3}
ALICE Collaboration, \href{https://cds.cern.ch/record/2803563/}{CERN-LHCC-2022-009, LHCC-I-038}.

\end{thebibliography}
\end{document}